\documentclass[12pt,a4paper]{article}

\let\section=\subsection     
\let\subsection=\subsubsection

\usepackage{amsmath}
\usepackage{amssymb}
\usepackage{mathrsfs}
\usepackage{times,mathptm}

\usepackage{afterpage}
\usepackage[dvips]{graphicx}

\newcommand{\ket}[1]{\ensuremath{ \big| \,{#1}\, \big> }}

\newcommand{\braket}[2]{\ensuremath{ \big< \,{#1}\, \big| \,{#2}\, \big> }}

\newcommand{\matrixe}[3]{\ensuremath{ \big< \,{#1}\, \big| \,{#2}\, 
\big| \,{#3}\, \big> }}

\newcommand{\partd}[2]{\ensuremath{ \frac{\partial #1}
{\partial #2} }}

\newcommand{\op}[1]{\ensuremath{%
    \fontdimen12\textfont3=2pt\fontdimen12\scriptfont3=1.4pt%
    \!\null\mathop{\vphantom{#1}\smash{#1}}\limits_{\!\sim}\null}\!}

\newcommand{\conj}[1]{\ensuremath{{{#1}}^{\star}}}
\newcommand{\hermit}[1]{\ensuremath{{{#1}}^{\dag}}}

\newcommand{\AMeV}{\ensuremath{A\,\mathrm{MeV}}}
\newcommand{\MeV}{\ensuremath{\mathrm{MeV}}}
\newcommand{\fm}{\ensuremath{\mathrm{fm}}}

\newcommand{\chemical}[2]{\ensuremath{{}^{#1}\mathrm{#2}}}

\newcommand{\calC}{\mathscr{C}}

\newcommand{\corr}[1]{\hat{#1}}

\newcommand{\cop}[1]{\op{\corr{#1}}}

\newcommand{\opC}{\op{C}}
\newcommand{\hopC}{\hermit{\opC}}
\newcommand{\opA}{\op{\mathcal{A}}}

\newcommand{\opH}{\op{H}}

\newcommand{\Rp}{R_{+}}
\newcommand{\Rm}{R_{-}}

\newcommand{\Rmx}{R_{-}(x)}

\newcommand{\Rmpx}{R_{-}'(x)}

\newcommand{\ketcQ}{\ket{\corr{Q}}}

\newcommand{\matrixeQ}[1]{\matrixe{Q}{#1}{Q}}

\newcommand{\braketcQ}{\braket{\corr{Q}}{\corr{Q}}}
\newcommand{\matrixecQ}[1]{\matrixe{\corr{Q}}{#1}{\corr{Q}}}

\newcommand{\Xx}{\vec{X},\vec{x}}

\begin{document}

\begin{center}
   {\large \bf REALISTIC INTERACTIONS AND CONFIGURATION}\\[2mm]
   {\large \bf MIXING IN FERMIONIC MOLECULAR DYNAMICS}\\[5mm]
    T.~NEFF, H.~FELDMEIER and R.~ROTH\\[5mm]
   {\small \it Gesellschaft f\"ur Schwerionenforschung mbH \\
   Planckstra\ss{}e 1, D-64291 Darmstadt, Germany \\[5mm] }
   J.~SCHNACK\\[5mm]
   {\small \it Universit\"at Osnabr\"uck, Fachbereich Physik \\
   Barbarastra\ss{}e 7,  D-49069 Osnabr\"uck, Germany \\[8mm]} 
\end{center}

\begin{abstract}

In Fermionic Molecular Dynamics the occurrence of multifragmentation
depends strongly on the intrinsic structure of the many-body state.
Slater determinants with narrow single-particle states and a cluster
substructure show multifragmentation in heavy-ion collisions while
those with broad wave functions, which resemble more a shell-model
picture, deexcite by particle emission. Which of the two type of states
occurs as the ground state minimum or as a local minimum in the energy 
depends on the effective interaction. Both may equally well reproduce 
binding energy and radii of nuclei. This ambiguity led us to 
reinvestigate the derivation of the effective interaction from realistic 
nucleon-nucleon potentials by means of a unitary correlation operator
which is much more suited for dynamical calculations than the G-matrix 
or the Jastrow method. First results of mixing many Slater determinants
are also presented.

\end{abstract}

\section{Fermionic Molecular Dynamics}

Fermionic Molecular Dynamics (FMD) \cite{fmd90,fmd95,fmd97} is a model
to describe ground states of atomic nuclei and heavy-ion reactions in
the low to medium energy regime below the threshold for particle
production.

The FMD trial state $\ketcQ$ takes care of the Pauli principle
explicitly by using a Slater determinant of gaussian single-particle
states $\ket{q_i}$.
\begin{equation}
  \ketcQ = \opC \opA \bigl( \ket{q_1}
    \otimes \cdots \otimes \ket{q_A} \bigr)
\end{equation}
$\opA$ is the antisymmetrization operator and $\opC$ is an optional
unitary correlation operator which will be discussed later.

The single-particle states $\ket{q_i}$ are gaussians with mean
position and mean momentum parametrized by 3 complex parameters
$\vec{b}$ and a dynamical complex width $a$. Spin $\ket{\chi}$ and
isospin $\ket{\xi}$ are usually parametrized as two-spinors
\begin{equation}
  \braket{\vec{x}}{q} =
      \braket{\vec{x}}{a, \vec{b}, \chi, \xi} = 
      \exp \bigl\{ - \frac{(\vec{x} -
        \vec{b})^2}{2 a} \} \ket{\chi} \otimes
      \ket{\xi} \: .
\end{equation}

The description with gaussian single-particle states is the closest
analogue to a classical phase-space trajectory and therefore allows
for a descriptive interpretation of the FMD time evolution.  The
gaussians form an overcomplete set and allow to represent shell-model
states as well as intrinsically deformed states.

The dynamical equations are derived from the time-dependent
variational principle
\begin{equation}
  \delta \int dt \frac{\matrixecQ{i \frac{d}{dt} -
            \opH}}{\braketcQ} = 0 \: .
\end{equation}
The variation with respect to the parameters $q_\nu$ which are
contained in the trial state $\ketcQ$ leads to the Euler-Lagrange
equations of motion
\begin{equation} 
  \label{eq:eom}
  i \sum_{\nu} \calC_{\mu\nu} \dot{q}_{\nu} = \partd{\mathscr{H}}{\conj{q}_{\mu}}
\end{equation}
with generalized forces given by the gradient of the Hamilton function
$\mathscr{H}$ and the matrix $\calC$ which describes the geometrical
structure of the fermion phase-space.
\begin{equation}
  \calC_{\mu\nu} = \partd{}{\conj{q}_{\mu}} \partd{}{q_{\nu}}
  \ln \braketcQ\: , \qquad 
  \mathscr{H} = \frac{\matrixecQ{\opH}}{\braketcQ}
\end{equation}
The initial state of a reaction, which is evolved in time according to
eq.~(\ref{eq:eom}), is the antisymmetrized product of boosted ground
states.

\section{Multifragmentation}

\begin{figure}[!b]
  \begin{center}
    \includegraphics[width=0.9\textwidth]{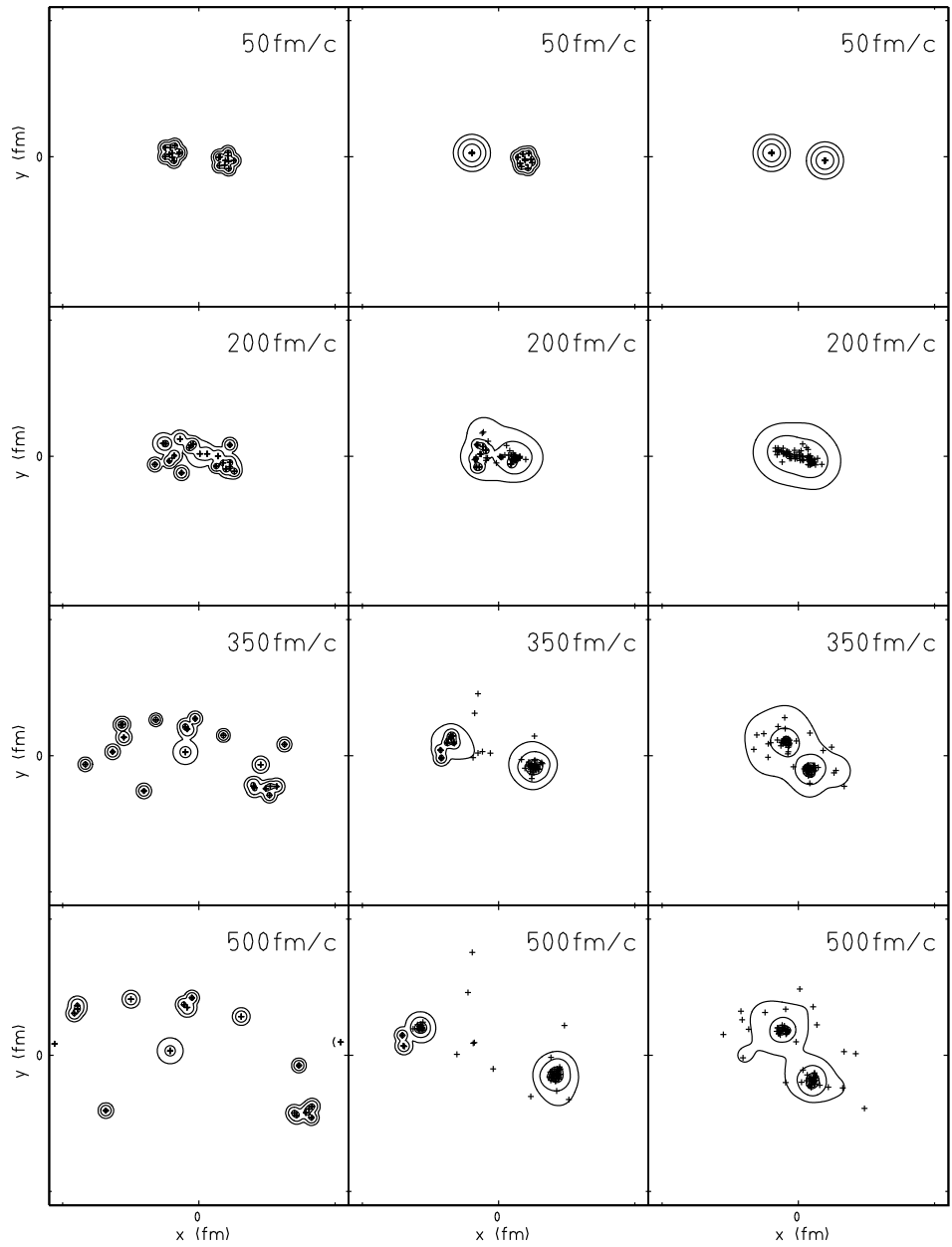}
    \caption{Density plots of $\chemical{40}{Ca}$+$\chemical{40}{Ca}$ at
      $E_{lab}=35$~\AMeV\/ and $b=2.75$~\fm. Crosses indicate
      centroids of gaussians.  Crosses without surrounding contours
      are from wave packets which have spread so much that their
      density is below the lowest contour (evaporated nucleon).}
    \label{fig:caca-reactions}
  \end{center}
\end{figure}

The results of FMD calculations for multifragmentation reactions show
a strong dependence on the intrinsic structure of the nuclear states
which is determined by the effective interaction.  All nucleon-nucleon
interactions which we use are adjusted to describe well binding
energies and radii of ground states.  They differ mainly in their
momentum dependent parts which are poorly determined by the ground
state properties but can lead to very different behavior in the
dynamics of a heavy-ion reaction. This effect is demonstrated in
fig.~\ref{fig:caca-reactions} where we display density contour plots
of $\chemical{40}{Ca}$+$\chemical{40}{Ca}$ reactions at energy
$E_{lab}=35$~\AMeV\/ and impact parameter $b=2.75$~\fm.

\afterpage{\clearpage}

The used phenomenological interaction has an FMD ground state with an
$\alpha$-cluster structure. Only 1~\MeV\/ above is a stationary FMD state
(local minimum) which shows no clustering in coordinate space but
looks more like a closed spherical sd-shell nucleus. These two
energetically almost degenerate states behave completely different in
heavy ion reactions. The clustered states lead to multifragmentation
where the spatial correlations in the initial state survive to a large
extend the collision. Reactions with the spherical states show no
multifragmentation. Here we observe binary inelastic collisions
followed by deexcitation through evaporation of single nucleons.
Collisions between the two different types of FMD states result in a
somewhat mixed situation.  It also happens that a smooth nucleus
sometimes jumps during a collision into a cluster configuration and
vice-versa.

\begin{figure}[b]
  \begin{center}
    \includegraphics[width=0.9\textwidth]{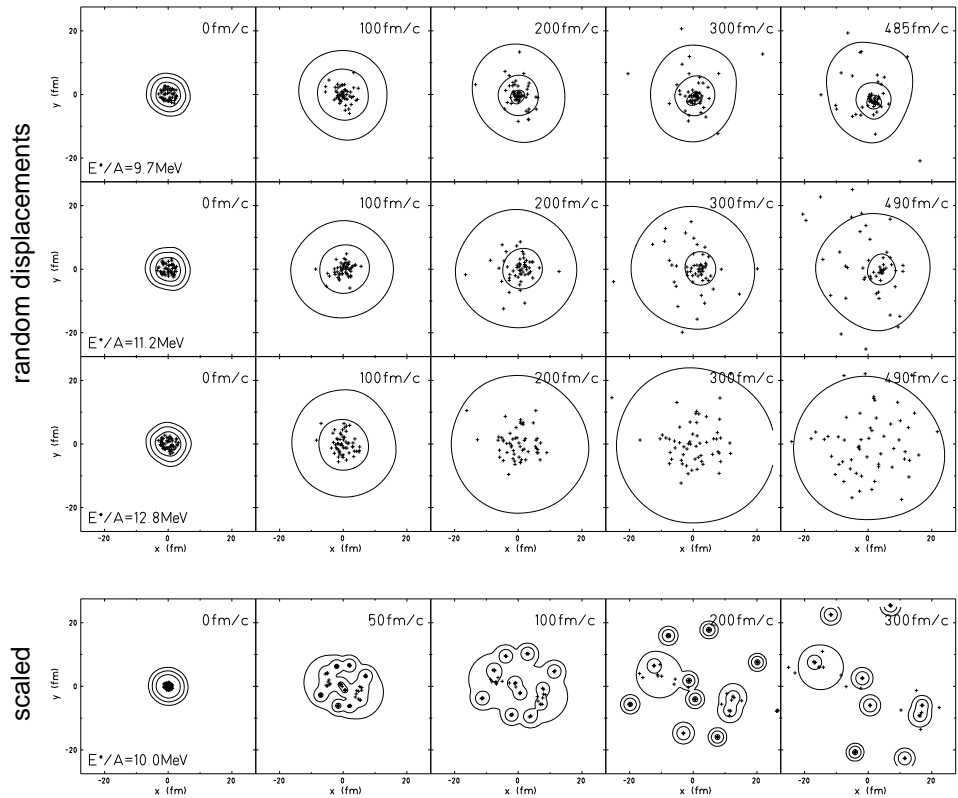}
    \caption{Density plots of the decay of excited $\chemical{56}{Fe}$
      nuclei.}
    \label{fig:fe-excitations}
  \end{center}
\end{figure}

The effect of initial correlations is further studied in the decay of
excited $\chemical{56}{Fe}$ nuclei where the initial excitation energy
was created in different ways.  In the upper rows of
fig.~\ref{fig:fe-excitations} we can see the effect of random
excitations which destroy the spatial correlations in the cluster
structure of the ground state.  The three different excitation
energies are achieved by randomly displacing the centroids of the
gaussian from their ground state positions but keeping the density at
normal values.  In all cases the nuclei show first an expansion caused
by the increased pressure.  If the excitation energy or the pressure
is not too high (first two rows) a contraction follows in the center
where the mean field is still strong enough to hold together a hot
fragment which finally deexcites by particle evaporation.  Typically
between 11 and 12 \AMeV\/ excitation the nuclei vapourize into
individual nucleons.

In the last row the excitation energy of 10 \AMeV\/ is created by
scaling down the distances between the centroids of the gaussians.  In
addition small random displacements are applied.  These excitations do
not destroy the spatial correlations between the nucleons and
multifragmentation into different clusters is observed.

The conclusion is, that in FMD initial correlations are important to
form clusters.  There is not enough time during the decay and
expansion of a randomly excited nuclear system to build up the
many-body correlations needed to form a rather cold fragment.  Either
the fragments originate from cool junks of the initial system or we
observe evaporation residues.

\section{Unitary Correlator Operator Method}

To avoid the above demonstrated ambiguities and to gain predictive
power we want to start from realistic nucleon-nucleon potentials.
There is however the old problem that realistic interactions, which
reproduce the scattering and deuteron data, feature a strong short
range repulsion and also tensor, spin-orbit and momentum-dependent
parts. The numerically convenient single determinant is only
appropriate if the system is dominated by a mean field but it is not
sufficient to represent the two-body correlations induced by the short
ranged repulsion and the tensor interaction.  Since we do dynamical
calculations with many time steps a G-matrix method with a
Pauli-operator which depends on the actual state and therefore on time
is not advisable.

Our approach to treat these correlations in a simpler fashion is
similar to the Jastrow method but in order to avoid complications with
a time-dependent norm in dynamical calculations we construct a unitary
correlation operator which does not depend on the actual time
evolution.  This conserves the norm of the correlated state and also
allows to apply the correlation operator in a state-independent way to
operators, resulting in correlated operators.

In a first step we developed the Unitary Correlation Operator Method
(UCOM) \cite{ucom98,roth:diplom,neff:diplom} to treat the strong
short-range repulsion of the realistic interactions.  The correlator
$\opC$ shifts the relative wave function of each pair of particles out
of the repulsive region of the interaction.  In two-body space it is
defined with the hermitian two-body operator $\op{S}$ which acts on
the relative coordinate $\vec{x}$ in the following way:
\begin{equation}
  \matrixe{\Xx}{\opC}{\Phi} =
      \matrixe{\Xx}{e^{- i\op{S}}}{\Phi} = 
      \exp \biggl\{ -\frac{1}{2} s'(x) - \frac{s(x)}{x} -
      s(x) \partd{}{x} \biggr\} \braket{\Xx}{\Phi} \: .
\end{equation}
$s(x)$ determines the amount by which the particles are shifted away
from each other.  Using correlation functions $\Rp$ and $\Rm$ defined
by
\begin{equation}
  \int_x^{\Rm(x)} \frac{dt}{{s}(t)} = -1\; , \qquad 
  \int_x^{\Rp(x)} \frac{dt}{{s}(t)} = +1 \: .
\end{equation}
We can write the correlated wave function in terms of a coordinate
transformation as
\begin{equation}
  \matrixe{\Xx}{\op{C}}{\Phi} =
  \frac{\Rmx \sqrt{\Rmpx}}{x} \braket{\vec{X}, \frac{\vec{x}}{x}
    \Rmx}{\Phi} \: .
\end{equation}

If we apply the correlation operator to operators we get the
corresponding correlated operators which act between uncorrelated
states.
\begin{equation}
  \matrixecQ{\op{B}} = \matrixeQ{\hopC \op{B} \opC} =
  \matrixeQ{\cop{B}} \:, \qquad \cop{B} = \hopC \op{B} \opC
\end{equation}

Of special interest is the correlated Hamilton operator $\cop{H}$.
Besides the transformed two-body potential $V(\Rp)$ we get two-body
interaction parts from the correlated kinetic energy which has
momentum-dependent and potential-like contributions.

Three-body and higher contributions from the correlated operators can
be neglected if the correlation volume times the density is small
enough so that the probability to find three ore more particles
simultaneously within the range of the strong repulsion is small.

\begin{figure}[!b]
  \begin{center}
    \includegraphics[width=0.95\textwidth]{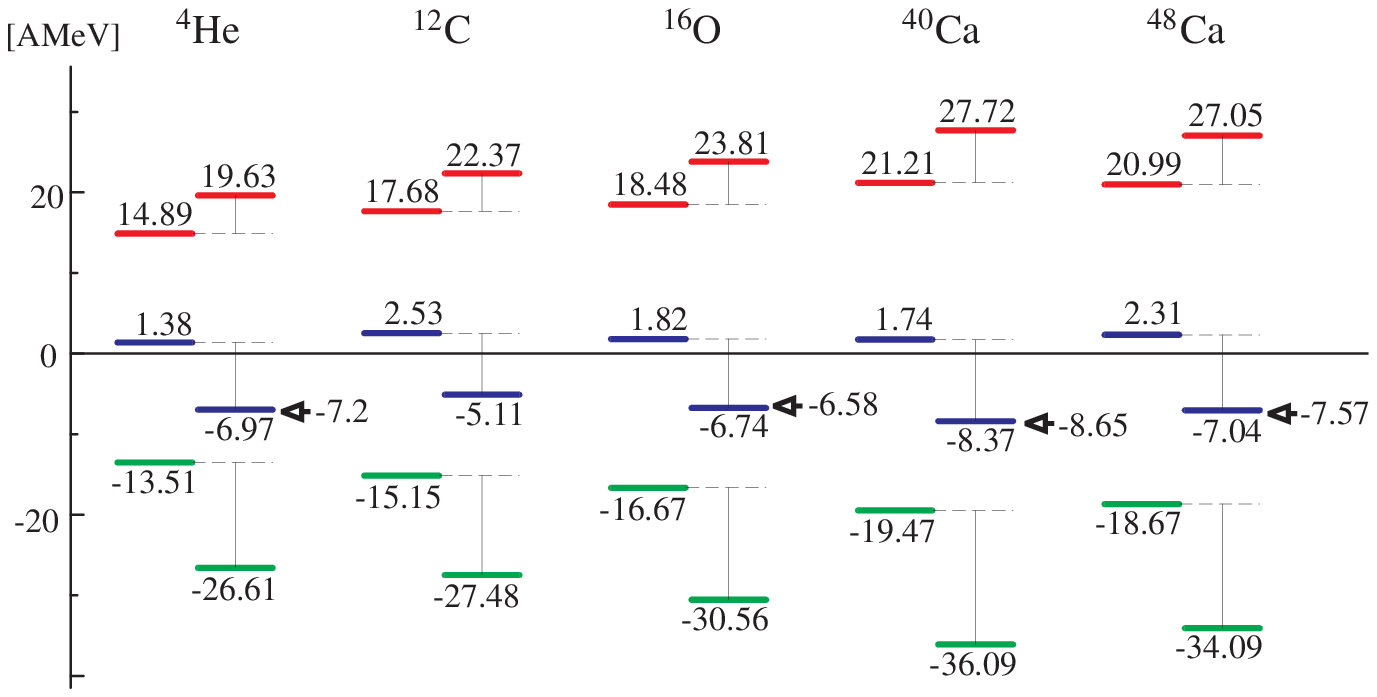}
    \label{fig:ats3m-energies}
    \caption{FMD calculations using the ATS 3M potential. For each
      nucleus expectation values of kinetic (top), potential (bottom) 
      and binding (middle) energy per nucleon are shown. 
      The left hand columns display the values for the uncorrelated and the 
      right hand columns for the correlated states.
      The arrows  indicate results of other methods, Yakubovski
      ($\chemical{4}{He}$), FHNC ($\chemical{16}{O}$) and CBF
      ($\chemical{40}{Ca}$ and $\chemical{48}{Ca}$). References
      in \cite{ucom98}.}

    \vspace{1em}

    \includegraphics[angle=0,width=0.48\textwidth]{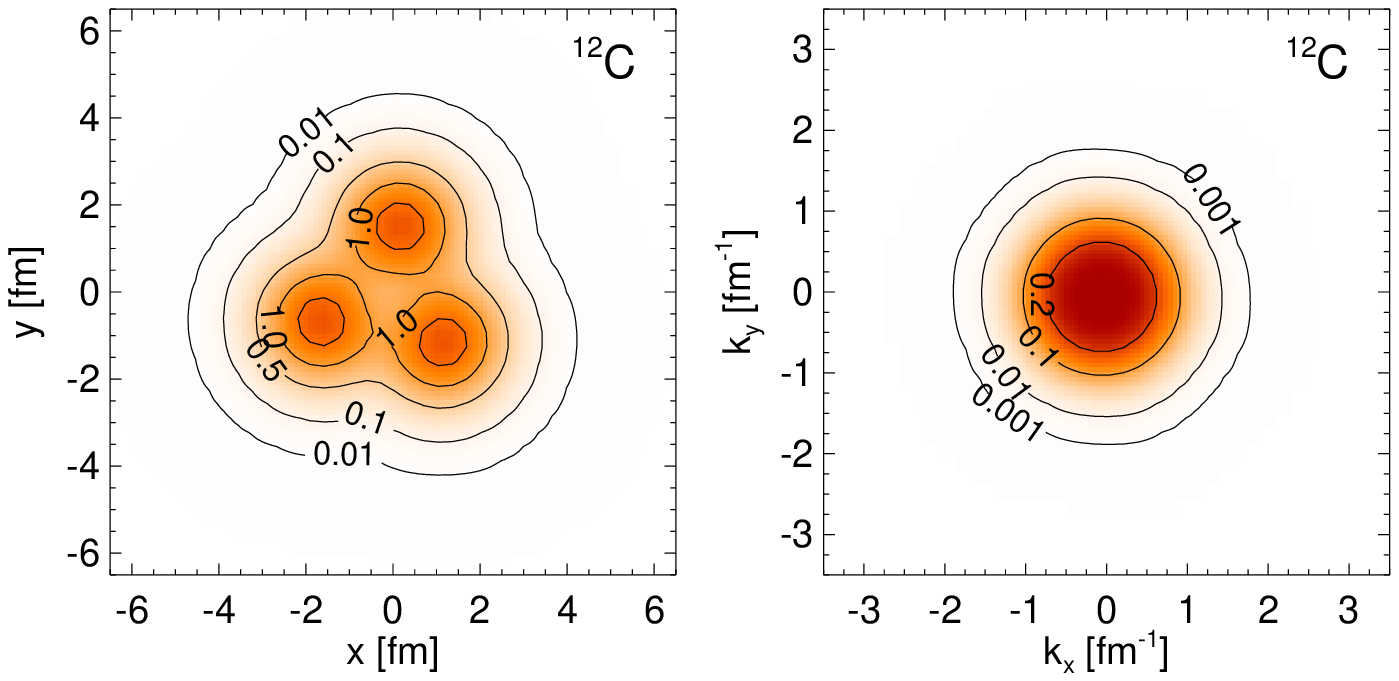}
    \includegraphics[angle=0,width=0.48\textwidth]{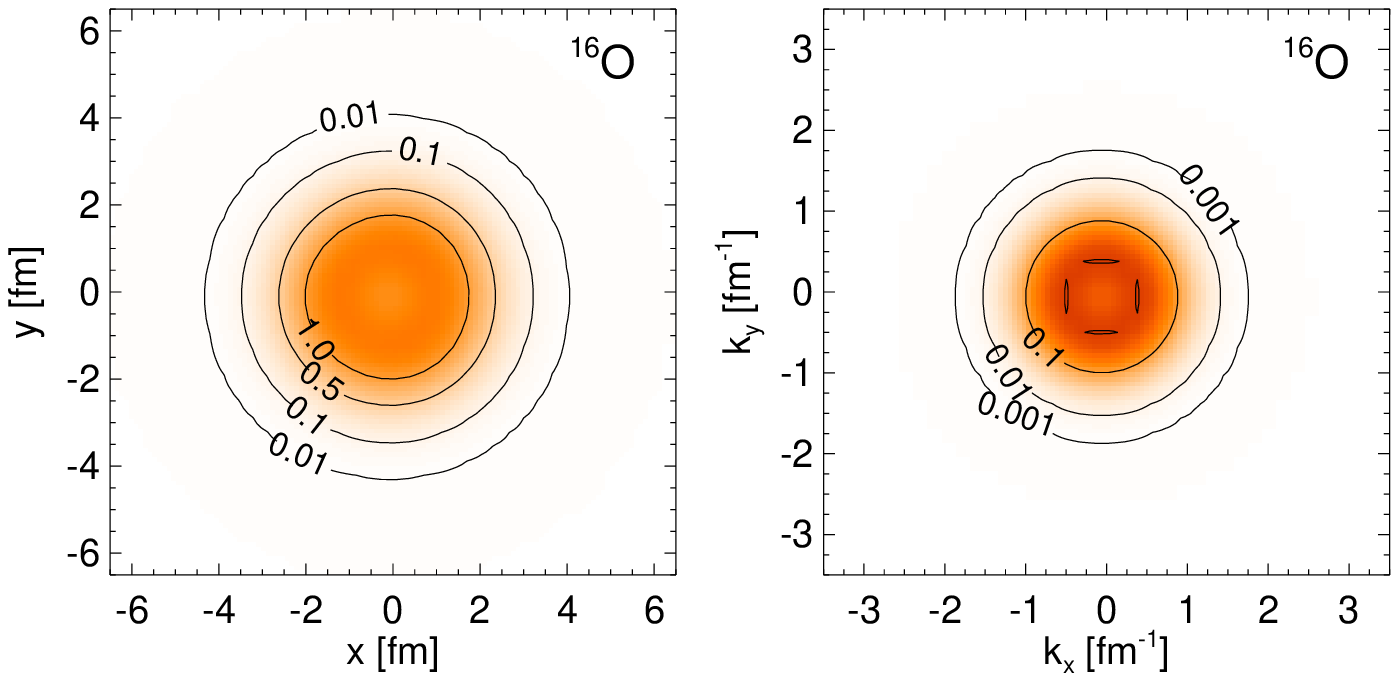}
    \includegraphics[angle=0,width=0.48\textwidth]{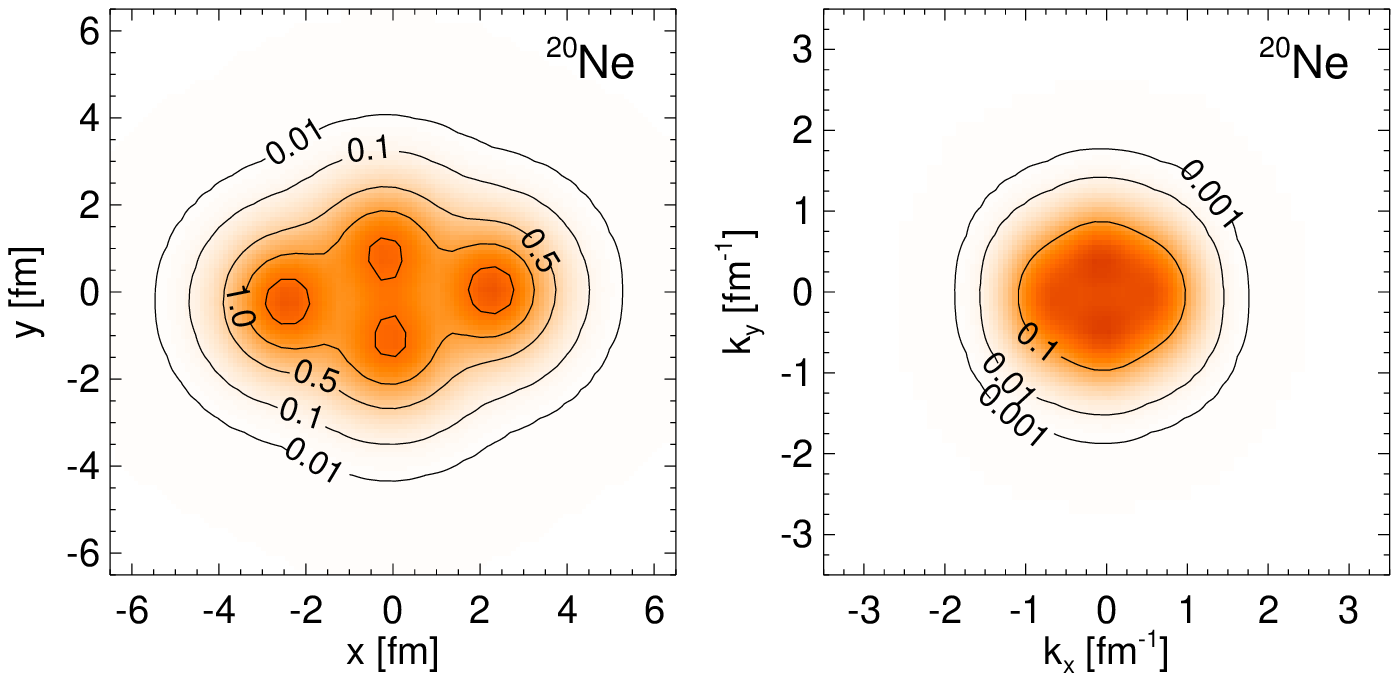}
    \includegraphics[angle=0,width=0.48\textwidth]{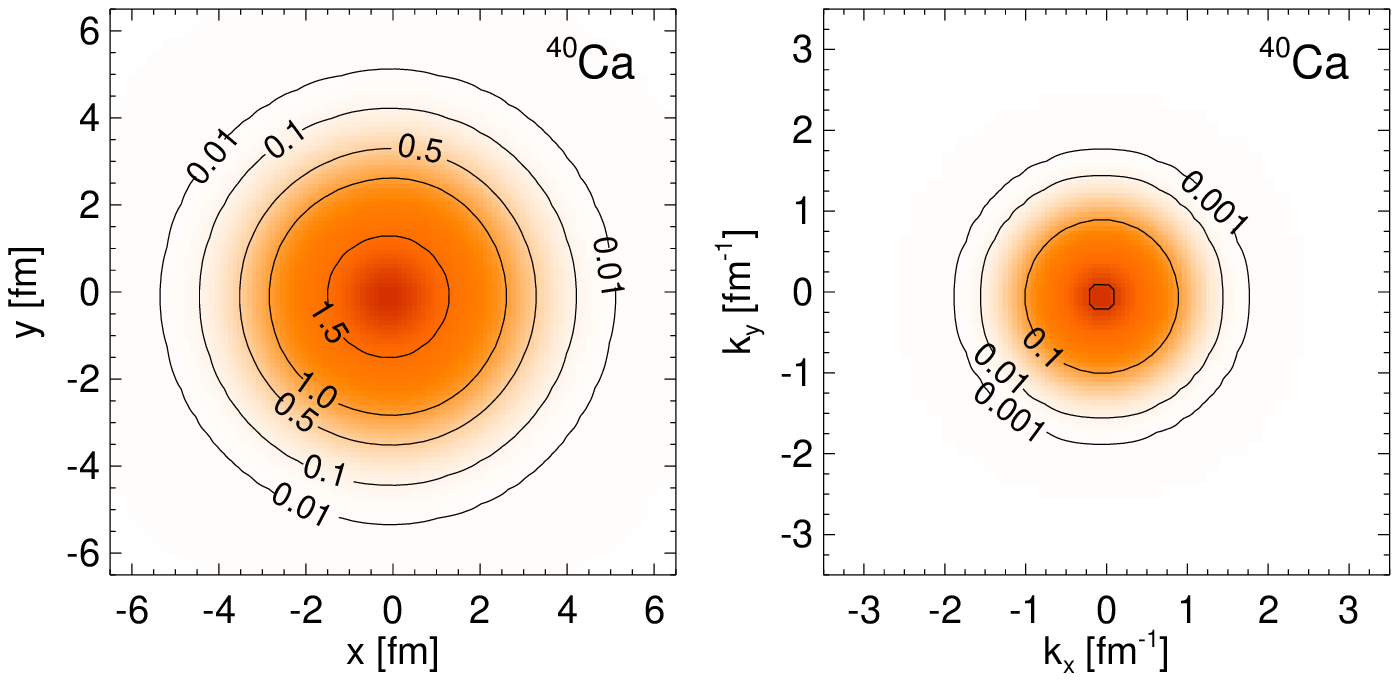}
    \caption{FMD ground states with the ATS 3M potential. Plotted are
      cuts of the nucleon density in coordinate and momentum
      space.}
    \label{fig:ats3m-densities}
  \end{center}
\end{figure}

As a test of the method we applied it to the Afnan-Tang S3M potential.
This pure central potential has been used as a benchmark for
many-particle methods. In fig.~\ref{fig:ats3m-energies} the results of
FMD calculations with the Unitary Correlation Operator Method are
shown. The agreement with other, numerically much more expensive,
methods is striking. The kinetic energy of the correlated state
increases in comparison to the uncorrelated one but this is
overcompensated by the gain in potential energy.  It is amazing to see
how accurately the large positive and large negative corrections from
the correlations add up to the correct binding energy.

Particularly interesting is the FMD result for $\chemical{12}{C}$.
Other methods have great difficulties to describe the intrinsic
structure of this nucleus well. The FMD result is shown in
fig.~\ref{fig:ats3m-densities}, where cuts of the nucleon density in
coordinate and momentum space are plotted. One can clearly see that
the FMD trial states with gaussians can describe both intrinsically
deformed states like $\chemical{12}{C}$ or $\chemical{20}{Ne}$ and
closed shell states like $\chemical{16}{O}$ or $\chemical{40}{Ca}$.

One should, however, keep in mind that in real nuclei a major part of
the binding originates from the tensor interaction which induces
correlations between the spin of two particles and the direction of
their relative distance vector. We are developing a unitary correlator
for these tensor correlations but the correlated hamiltonian is of a
rather complicated form and results are not available yet.

\section{Configuration Mixing}

If one wants to address questions of nuclear structure in the FMD
environment more refined trial states are necessary and possible. The
parameterization of the one-particle state can be improved by using a
superposition of several gaussians. This strategy promises to be
useful for the description of halo-nuclei with their far out reaching
exponential tail in the nucleon density \cite{neff:diplom}.

\begin{figure}[htbp]
  \begin{center}
    \includegraphics[width=0.7\textwidth]{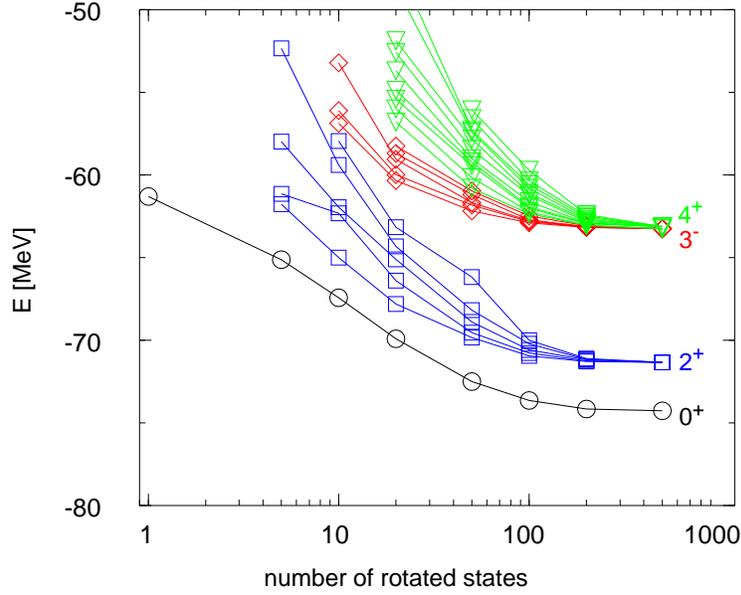}
    \caption{Results of configuration mixing calculations -- plotted 
      are the energies of the lowest states as a function of the basis 
      dimension.}
    \label{fig:c12-rotated}
  \end{center}
\end{figure}

On the other hand superpositions of Slater determinants can lead to a
better description of medium and long ranged many-particle
correlations. To demonstrate this approach we present an improved
treatment of the $\chemical{12}{C}$ ground state. As shown in the last
section the FMD ground state is given by an intrinsically deformed
single Slater determinant which of course lacks the symmetries of the
real $\chemical{12}{C}$ state regarding parity and angular momentum.
As an alternative to the projection on the right quantum numbers we
perform a configuration mixing calculation in a set of randomly
rotated FMD states.  Formally this leads to a generalized eigenvalue
problem where the Hamilton operator is represented in a nonorthogonal
set of FMD states $\bigl\{ \ket{Q^i} \bigr\}$:
\begin{equation}
  \sum_j \matrixe{\corr{Q}^i}{\op{H}}{\corr{Q}^j} \; c^{\alpha}_{j} =
    E^{\alpha} \sum_j \braket{\corr{Q}^i}{\corr{Q}^j} \;
    c^{\alpha}_{j} \: .
\end{equation}
The energies of the lowest eigenstates of such a configuration mixing
calculation are shown in fig.~\ref{fig:c12-rotated} as a function of
the number of basis states. With increasing number of basis states the
lowest states become better and better eigenstates of parity and
angular momentum and the rotational bands emerge.  One can also
observe a rather large increase in binding energy of about 12~\MeV\/ for
the ground state.

\section{Some Ideas about Quantum Branching}

Since the antisymmetrized products of gaussians form an overcomplete
basis set in Fock space any many-body state, also the exact solution
of the Schr\"odinger equation, can in principle be represented by a
superposition of FMD states.  The configuration mixing calculation for
$\chemical{12}{C}$ in the previous section shows for example the
ground state as a superposition of many Slater determinants.  In a
dynamical calculation we can for numerical reasons only follow the
evolution of a single component.  But like in the stationary situation
this component mixes via off-diagonal matrix elements of the
hamiltonian with other determinants during the time evolution. Many
models, like AMD \cite{amdv96,amd:hirschegg99} or QMD
\cite{qmd:hirschegg99} simulate these quantum branchings by means of
random collision terms.

A more refined treatment of this effect should be possible by allowing
the FMD state $\ket{Q(t)}$ to have a certain possibility to jump to
another FMD state $\ket{Q'(t)}$. This branching to another trajectory
should be determined by the perturbation operator
\begin{equation}
  i \sum_{\nu} \dot{q}_{\nu} \partd{}{q_{\nu}} - \opH
\end{equation}
which describes the difference between the FMD and the exact
time-evolution.  Open problems are the conservation laws and the
approximation needed to come from perturbative transition amplitudes to
transition probabilities.  If the system is in an energy regime with
high level density statistical arguments may be employed.

Quantum branching is probably not only needed in multifragmentation to
jump from a situation with wide wave packets to cluster states, but
also in general to allow for example crossing of potential barriers
which exist in the highly restricted phase space of the parameters but
can be tunneled in reality.

Another example is the breaking of symmetries.  The exact final state
possesses the dynamically conserved symmetries of the initial state
simply by a superposition of a channel and its counterpart.  An
illustrative example is the left-right mirror symmetry of the
$\chemical{40}{Ca}$+$\chemical{40}{Ca}$ reaction.
After the collision the measured channels are of course not symmetric, 
only the superposition of all channels has the proper symmetry.
When in the approximate scheme only a single state,
which has initially the mirror symmetry, is evolved in time,
it will conserve this symmetry in a to restricted way.
During the evolution the symmetry should be broken by 
quantum branching such that with equal probability each final
channel and its mirrored counterpart can be reached.

\bibliographystyle{fmdplain}
\bibliography{fmdbib}

\end{document}